# A Probabilistic Framework for Quantifying Biological Complexity

Stuart M. Marshall, Alastair R. G. Murray and Leroy Cronin*

*School of Chemistry, University of Glasgow, Glasgow, G12 8QQ, UK. *Corresponding author email: Lee.Cronin@glasgow.ac.uk*

**Abstract**

One thing that discriminates living things from inanimate matter is their ability to generate similarly complex or non-random architectures in a large abundance. From DNA sequences to folded protein structures, living cells, microbial communities and multicellular structures, the material configurations in biology can easily be distinguished from non-living material assemblies. This is also true of the products of complex organisms that can themselves construct complex tools, machines, and artefacts. Whilst these objects are not living, they cannot randomly form, as they are the product of a biological organism and hence are either technological or cultural biosignatures. The problem is that it is not obvious how it might be possible to generalise an approach that aims to evaluate complex objects as possible biosignatures. However, if it was possible such a self-contained approach could be useful to explore the cosmos for new life forms. This would require us to prove rigorously that a given artefact is too complex to have formed by chance. In this paper, we present a new type of complexity measure, Pathway Complexity, that allows us to not only threshold the abiotic-biotic divide, but to demonstrate a probabilistic approach based upon object abundance and complexity which can be used to unambiguously assign complex objects as biosignatures. We hope that this approach not only opens up the search for biosignatures beyond earth, but allow us to explore earth for new types of biology, as well as observing when a complex chemical system discovered in the laboratory could be considered alive.

**Keywords** complexity; biosignature; pathway complexity; living-non-living threshold

**Introduction**

**Biosignatures**

There have been many proposals for finding effective biosignatures, that is, unambiguous indicators of the influence of life in an environment. These include searching for atmospheric gases such as methane [1], looking for signs of a distinctive $^{56/54}$Fe isotope ratio [2], searching for biological impact on minerals and mineral assemblages [3], for fossils [4], or for distinctive patterns in the distribution of monomer abundance [5]. It has also been suggested that life on exoplanets could be detected by searching for a variant of the distinctive "red-edge" spectral feature of the Earth [6], where there is a strong increase in reflectance in the 700 to 750 nm region of the spectrum due to vegetation. In this case, we cannot necessarily expect alien vegetation to share the spectral characteristics of terrestrial vegetation, but perhaps a spectroscopic signature could be observed at another wavelength. Additionally, extreme care should be taken to avoid misidentifying this effect with similar effects that can be caused by certain mineral formations. These two caveats highlight particular difficulties in trying to classify phenomena as biosignatures. The first difficulty is in ensuring that we cast the net wide enough to include biologies that may well differ fundamentally from our own. By remaining too tied to the details of terrestrial biology, we risk missing biosignatures presented to us due to our assumptions about what life must be like. The second difficulty is in avoiding false positives by ensuring abiotic causes are ruled out. For example, shortly after the $^{56/54}$Fe ratio was suggested as a biosignature in 1999 [2], Bullen et al published in 2001 evidence that the same isotopic fragmentation could have abiotic origin [7]. In another example, a 2002 paper declared that magnetite crystals within Martian meteorite ALH84001 were "A Robust Biosignature"[8], however potential abiotic processes to create such crystals have also since been proposed [9][10].

**Complexity Measures**

The concept of complexity is itself curious since even discussion about its nature is complicated. This is because there is currently no consensus on a single unambiguous definition [11]. In addition, descriptions of complexity and randomness are intrinsically related and many definitions of complexity are specific to certain fields or applications, as well as needing an often-biased observer which can end up comparing intrinsically different things. This is a problem since it can result in misleading notions about which object is more complex. We will describe below some existing measures of complexity, although as there are a great many such measures the list is far from exhaustive. Several complexity measures find utility in the realms of computation and information. In information theory, the "Shannon Entropy" [12] of a string of unknown characters, which can be used as a complexity measure, is a measure of how predictable the outcome of the string is, or equivalently how much information it contains, based on the probability of each possible character in the string. The Kolmogorov complexity of a known object [13] is the minimum length of program that outputs the object, where for example strings containing many repetitions would have lower complexity than those that are random. Logical depth [14] is a complexity measure somewhat similar to Kolmogorov complexity, but looking at the time required to generate the object from a random input, rather than the size of the program. Effective complexity [15] looks for a compressed description of the regularities of an object. One can also examine the computational complexity [16] of an algorithm, which gives a measure of how the resources required increase with the size of the input. Stochastic complexity is another similar measure, but which looks at the shortest encoding of the object taking advantage of probabilistic models [17]. Measures such as Shannon entropy and Kolmogorov complexity are maximum for random structures, although one can argue that randomness is not necessarily complexity, and that maximum complexity lies somewhere between completely ordered and random structures [11].

There have been a number of suggestions for complexity measures on molecules [18], or crystal structures [19]. These range from those based on information theoretic measures, to specific features of the chemical graph such as vertex degrees, [20] and the number of subgraphs of the molecular graph [21]. In other applications, measures for complexity in graph theory [22], in tile self-assembly [23], and in biology in relation to genes and their environment have been proposed [24]. Here we present the concept of Pathway Complexity which identifies the shortest pathway to assemble a given object by allowing the object to be dissected into a set of basic building units, and rebuilding the object using those units. Thus the Pathway Complexity can be seen as a way to rank the relative complexity of objects made up of the same building units on the basis of the pathway, exploiting the combinatorial nature of these combinations, see Figure 1.

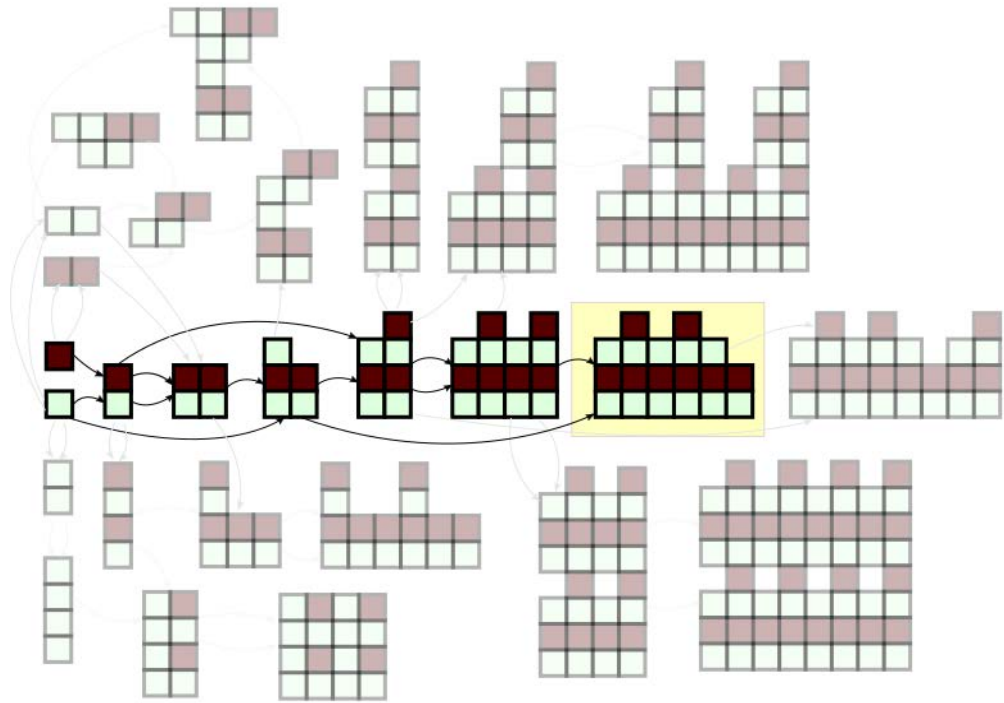

*Figure 1: Illustration of a complexity pathway in blocks, with the target shown by the yellow box. A combinatorial explosion in structures is illustrated by the other faded structures shown, which are just a small set of the many alternative structures that could be constructed.*

**Results and Discussion**

**Pathway Complexity as a Biosignature**

We propose a measure of complexity based on the construction of an object through joining operations, starting with a set of connected substructures, where structures already built in the process can be used in subsequent joining operations. The sequence of joining operations that constructs the objects can be defined as a complexity pathway, and the number of associated joining operations is defined as the complexity of that pathway. The complexity of the object in relation to the set of substructures is defined as the lowest complexity of any pathway. We call this complexity measure 'Pathway Complexity' and it is illustrated in Figure 2.

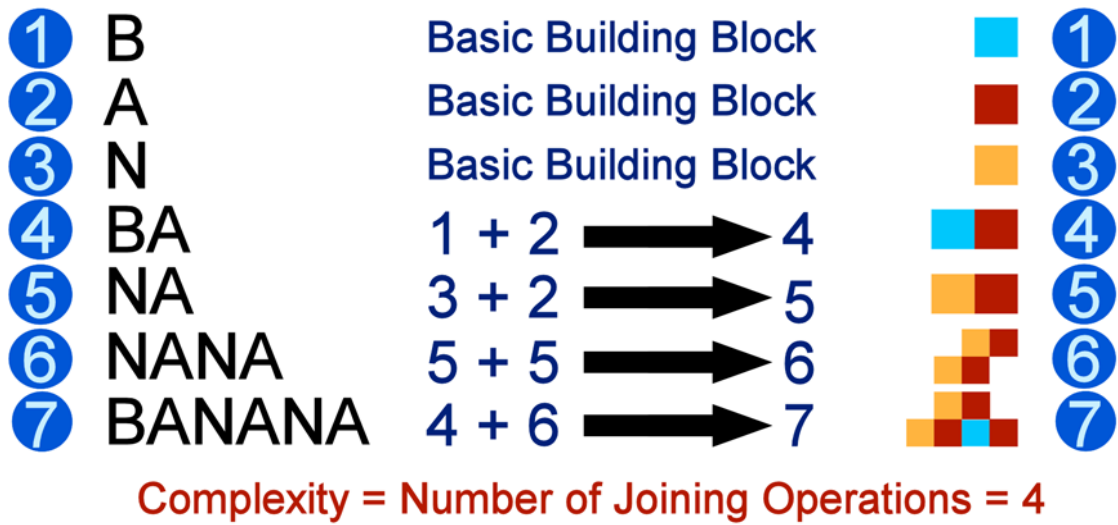

*Figure 2: Complexity pathways for a text string and a simple block shape giving the Pathway Complexity to construct the word Banana from its basic building blocks as 4.*

The motivation for the formulation of Pathway Complexity is to place a lower bound on the likelihood that a population of identical objects could have formed abiotically from an initial pool of starting materials, i.e. *without the influence of any biological system or biologically derived agent*. An object of sufficient complexity, if formed naturally, would have its formation competing against a combinatorial explosion of all other possible structures. If that any given object was found in abundance, it would be a clear indication that life-like processes were required to navigate in the state space to that particular structure, rather than diffusing the starting materials through the state space and ending up with a diverse mixture of structures that may or may not contain the structure in question. This *non-trivial trajectory in the state space* is, we propose, a characteristic unique to living systems. Therefore, if we can use Pathway Complexity to place a lower bound on the threshold where a trajectory becomes non-trivial, we can then establish whether an object is undoubtedly of biological origin. By following this reasoning it can be proposed that living systems themselves are self-sustaining non-trivial trajectories in a state space. This means that the biosignatures produced by living systems are themselves non-trivial trajectories [25]. As such, Pathway Complexity bounds the likelihood of natural occurrence by modelling a naïve synthesis of the object from populations of its basic parts, where at any time pairs of existing objects can join in a single step. In establishing the Pathway Complexity we are asking, in this idealised world, if the number of joins required would be low enough that we could have some population of the desired object rather than being overwhelmed by instances of all the other structures that could be created. Of course, some pathways may be more favoured than others (such as in chemical synthesis), but unless we have special pathways with 100% yield of each substructure on the pathway, then that fact merely pushes back the threshold. If we find anything significantly above the threshold then this, we propose, is a general biosignature. By searching for complexity alone, whether of molecules, objects, or signals, we don't have to make any assumptions about the details of the biology or its relation to our own biology. This approach therefore offers a new approach, and we show below that a rigorous framework can be developed to search for agnostic biosignatures.

**Pathway Complexity: Basic Approach**

The basic approach for determining the Pathway Complexity of an object is applicable when we are considering the construction of an object in its entirety from defined, basic subunits. The Pathway Complexity is calculated in the context of any possible objects that could be constructed from the same subunits. Later we will extend this approach to assessing the complexity of a class of objects that are not necessarily identical. We represent subunits of the object as vertices, and connections as edges, in a graph $G$. The vertices of $G$ are grouped into equivalence classes, which in the basic approach would mean that subunits in the same equivalence class are identical. There may be multiple types of edges if there are different types of connection in the object. We then construct complexity pathways for $G$ and establish their complexities using the following process. We start with a sequence containing only trivial "fundamental" graphs representing each unique subunit in the object. A pair of graphs in the sequence is joined by adding one or more edges between vertices of one graph and the other, see Figure 3. A pathway is complete when the sequence contains $G$, i.e. when the graph of the object has been constructed. The complexity of the pathway is the number of joining operations required to complete the pathway. The Pathway Complexity of $G$, and of the object with respect to the given substructures, is the smallest number of joining operations of any pathway. $G$ may be a directed graph, if the direction of a connection is important. For example, in a text string, different structures will result in connections left-to-right and right-to-left.

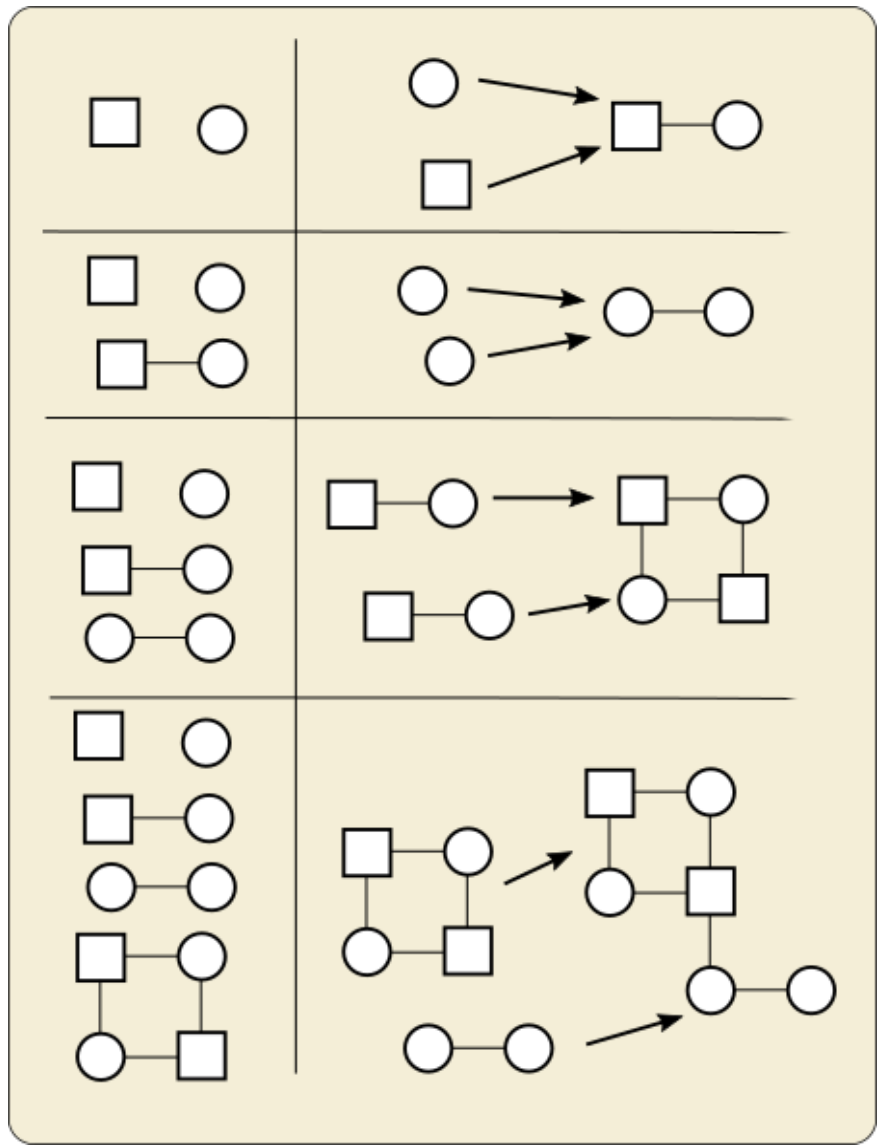

*Figure 3: Illustration of a single complexity pathway, with the set of objects on the left and the joining operations on the right. At each step, the structure created by the operation is added to the set of objects and available for subsequent joining operations.*

We can describe a search tree representing all different pathways, as at each point allowable combinations of different graphs in the set, with different edge types, joined at different combinations of vertices, would follow a pathway down a different branch of the tree, provided the graph resulting from the join is an induced subgraph of $G$. It should be noted that while we are conceptually exploring the entire search tree, in practice it is not necessary to explore every pathway as described above, as algorithmic implementations including branch and bound, and other techniques, can reduce the computational burden.

**Choice of substructures**

The choice of the basic substructures depends on the context of the desired complexity. For example, if we are establishing the complexity of the word "banana" then we could select the set of unique structures $\{b, a, n\}$, where the complexity is relative to all other words that can be made from those three letters. In fact, we could reasonably extend this set to include all letters of the English language plus punctuation, so we could then compare the complexities of any arbitrary phrase in any language using that alphabet. For a chessboard, natural units to choose would be {black square, white square}, and the complexity would then be in relation to all patterns that can be made of black and white squares.

In selecting the set of basic subunits we need to consider the class of objects that we are comparing. For example, if comparing a polymer to all polymers made of the same types of monomer, then the monomers could be our basic subunits, but if being compared to all molecules in general then we would be likely to select atoms types or bond types instead.

**Mathematical Formulation for the Pathway Complexity of graphs**

The following is a mathematical formulation for establishing the complexity pathway of a graph, as described above.

**Definition 1.** A graph $G$ can be **constructed in one step** from two graphs $X$ and $Y$ iff:

- X and Y are disjoint subgraphs of G
- Every vertex in $G$ is in either $X$ or $Y$
- Every edge in $G$ is either in $X$, in $Y$, or connects a vertex in $X$ with a vertex in $Y$

**Definition 2.** A **Complexity Pathway** of a graph $G$ relative to a set of m single-vertex graphs is defined as a sequence of graphs $G_{-m+1}, G_{-m+2}, \ldots, G_0, G_1, G_2, \ldots, G_n$ such that:

- $G_n = G$
- for $i < 1$, $G_i$ is a single-vertex graph

- for $i \geq 1$, $G_i$ can be constructed in one step from two graphs $G_j$ and $G_k$, with $j, k < i$

**Definition 3.** The **Pathway Complexity** $C$ of $G$ is the length of the shortest complexity pathway of $G$, minus the number of single-vertex graphs in that pathway (i.e. *n* in Definition 2). In other words, $C$ is the smallest number of construction steps, as defined in Definition 1, that will result in a set containing $G$.

**Characteristics of Pathway Complexity**

The Pathway Complexity of an object generally increases with size, but decreases with symmetry so large objects with repeating substructures may have lower complexity than smaller objects with greater heterogeneity. In addition, the history dependence and recursive nature of the measure means that internal symmetries are also accounted for if they lie on the shortest pathway. For example, an object may be asymmetric but have a symmetric feature in it that can be constructed through duplication prior to the asymmetric parts being added on. Those duplicated structures may themselves contain substructures with similar duplications, which are accounted for recursively. In this way, we can describe the construction of structures through repeated duplication and addition of subunits.

Pathway Complexity has an upper bound of $N_v - 1$, where $N_v$ is the number of vertices on the graph. This represents joining two fundamental graphs in the first step, and then adding one more at a time until the object is constructed. One lower bound of Pathway Complexity is $\log_2 N_v$, which represents the fact that the simplest way to increase the size of an object in a pathway is to take the largest object so far and join it to itself, e.g. we can make an object of size 2 with one join, 4 with 2 joins, 8 with 3 joins, etc. An illustration of the upper and lower bounds of Pathway Complexity can be seen in Figure 4, with the orange regions being forbidden due to the above boundary conditions. The green portion of the figure is illustrative of the location in the complexity space where life might be reasonably be found. Regions below can be thought of as being potentially naturally occurring, and regions above being so complex that even living systems might have been unlikely to create them. This is because they represent structures with limited internal structure and symmetries, which would require vast amounts of effort to faithfully reproduce. In exploring this region, we can attempt to find these boundaries, and examine the rate at which living systems can increase their complexity, and the limitations on that increase.

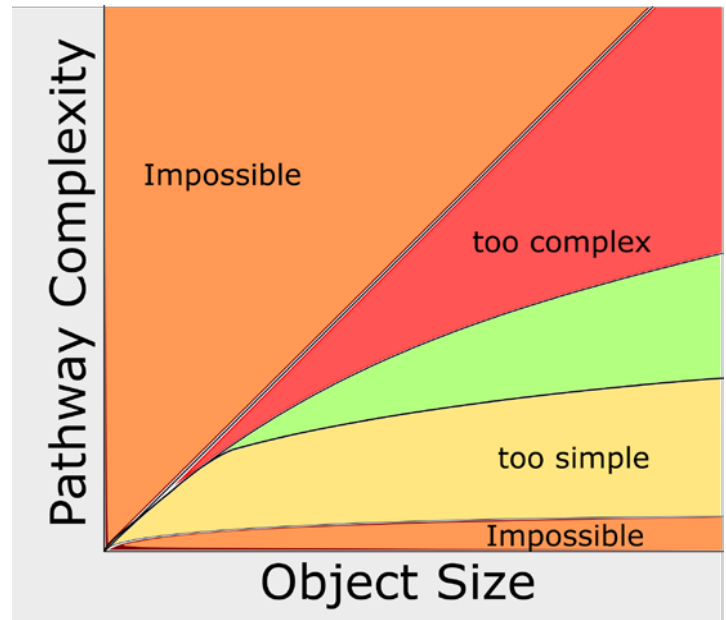

*Figure 4: An illustrative graph of complexity against size of the state space. Orange regions are impossible as they are above or below the bounds of the measure. The green region is where living systems may be most likely, where structures are neither too simple to be definitively biological, nor too complex to exist at all.*

**Example - Text**

Pathway Complexity can be used to examine text strings, finding the shortest complexity pathway by leveraging internal regularities. In the following example, we used an algorithm to analyse four strings of text to establish their Pathway Complexity. The following text strings were used, each of them 60 characters long. For simplicity, we have converted the strings to lower case without space or punctuation.

1) A random sequence of letters:
   "anpncsaveuoaklkgobqfdfoqtyilrzausbcbsxfclanbipcwizlmajbualbs"

2) Some text from "Green Eggs and Ham"[26] by Dr. Seuss:

"iamsamiamsamsamiamthatsamiamthatsamiamidonotlikethatsamiamdo"

3) Some text from "Dracula"[27] by Bram Stoker:

"myfriendwelcometothecarpathiansiamanxiouslyexpectingyousleep"

4) A highly repetitive sequence:

"redrumredrumredrumredrumredrumredrumredrumredrumredrumredrum"

Intuitively, one would expect the ascending order of complexity to be 4, 2, 3, 1 (with 2 simpler than 3 as "Green Eggs and Ham" is known for its simplicity and repetition). This ordering was confirmed by the algorithm, which found the Pathway Complexity of the (4) to be 9, of (2) to be 26, of (3) to be 53, and of (1) to be 57.

The maximum possible Pathway Complexity for any 60-character sequence is 59, so we would expect a random string to have a value close to this. In our random string (1), there are two repetitions in the sequence that the pathway has leveraged to reduce the complexity to 57, which are repetitions of "an" and "bs".

In (3), the passage from Dracula, the pathway found has used repetition of "ec", "ia", "an", "th", and "ousl".

In (2), the algorithm constructs "am" and then uses that in "sam" and "iam". It then constructs "samiam" from that pair, and adds letters "t", "h", "a", "t", to make "thatsamiam". These are then used in the final pathway where "samiam" and "thatsamiam" are repeated two and three times respectively.

The algorithm in (4) constructs the phrase "redrum" from individual letters, and then duplicates that to make "redrumredrum", further duplicating that to make "redrumredrumredrumredrum", then "redrumredrumredrumredrumredrumredrumredrum-redrum". Finally, "redrumredrum" is added to give the result.

**Complexity: General Approach**

We can extend the basic complexity measure above to cope with assessing the complexity of a group of objects that contain identical connection motifs. In this case we examine a population of objects and abstract out a common graph based on connected subunits that share features. For example, if examining a set of cups or mugs we can create a common graph of "handle connected to body", regardless of potential variations in size/colour etc. If examining a set of human beings, then we could create a common graph of bone connectivity, ignoring variations in size/shape of individual bones, or any material in the body other than bones.

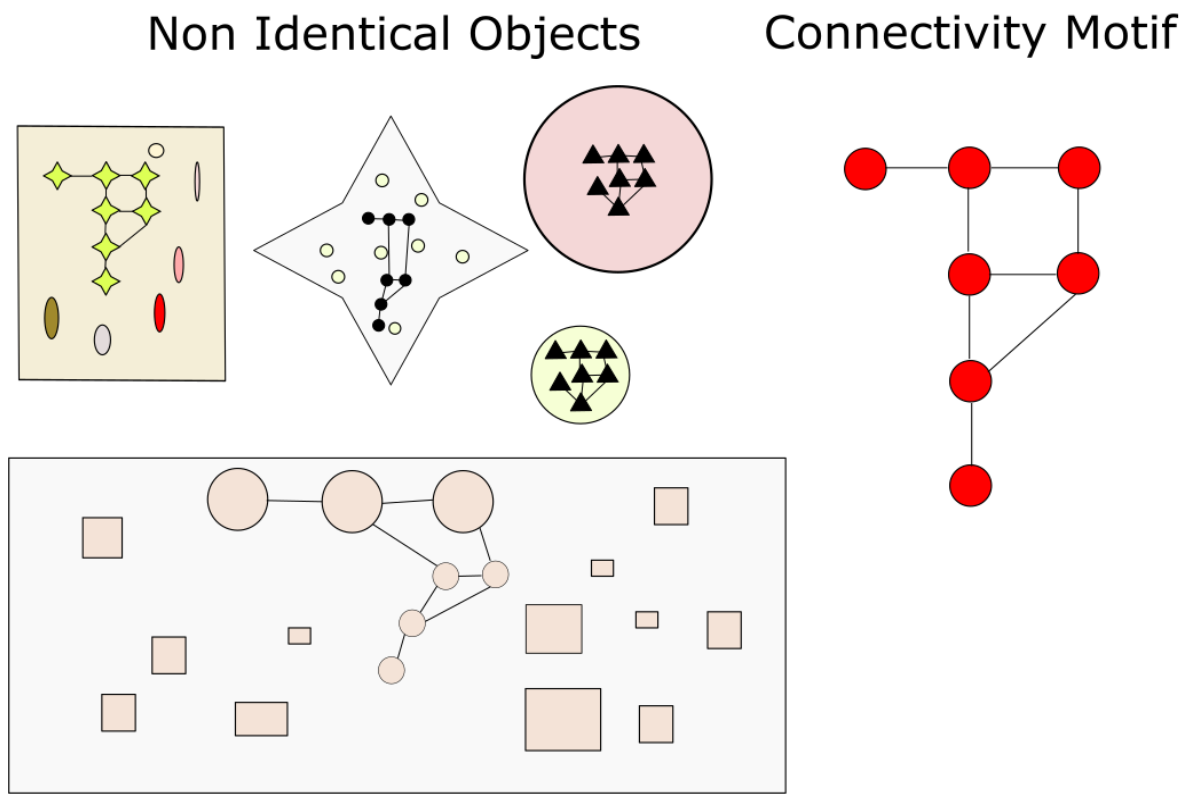

*Figure 5: Illustration of the general approach. The same connection motif can be found in all of these shapes, as shown in the graph on the right. Even though the structures and their components are quite different, we can extract the same graph from them and establish its Pathway Complexity.*

**Choice of subunits and connections**

In the general case, we define an archetypal set of connected subunits $S = \{s_1, s_2, s_3, ...\}$, along with a set of equivalence classes for the subunits $P = \{p_1, p_2, p_3, ...\}$, and a function $f: S \rightarrow P$. $f$ maps members of $S$ into equivalence classes in $P$ based on defining characteristics for each of the $p_i$. For example, looking at bones in the human body there could just be a single class for "bone" with the characteristic "made of bone", or different classes for different types of bone distinguished by some characteristic of that type of bone (e.g. tibia, sternum). There may be characteristics of members of $S$ not considered by $f$ and these can be thought of as "noise". For example, the same type of bone will vary in shape and size across individuals, but we are only interested in the characteristics that define the mapping onto $P$. Connections are defined by mapping into another set of equivalence classes $E = \{0, e_1, e_2, \dots\}$ by some function $g: S \times S \to E$. $E$ contains the 0 element to represent "not connected", and the $e_i$ represent different types of connection. Here connections could be actual physical connection, or it could be some more abstract relationship. We then define the archetypal graph $G$, in which vertices are members of $S$, with categories $f(s_i)$, and an edge exists between $s_i$ and $s_j$ if $g(s_i, s_j) \neq 0$, with edges categorised by $g(s_i, s_j)$. In the general case, we are looking at a class of objects to which the above rules can be applied to extract a graph isomorphic to $G$. In this case, members of $S$ are not necessarily substructures that can rebuild an object in its entirety, but rather are shared connection motifs common to a number of objects that we consider to be similar/related.

In the construction of $G$, it is important that for each member of $P$ all subunits that are common to objects in the class being examined, and that any shared instances of substructures within P in the objects are included in $S$. This is to prevent overestimation of complexity by selecting a more complex subgraph of $G$ through exclusion of some member of $S$. For example, one could remove some internal symmetries of a skeleton by selectively erasing some of the bones.

**Pathway Complexity in the general case**

The procedure for constructing complexity pathways on $G$, and defining the Pathway Complexity of the object, then follows that of the basic approach, only now we are establishing the Pathway Complexity of the selected archetypal graph $G$ that is contained within the whole class of objects. In this way, we can bound the complexity of sets of object that are non-identical but that clearly share features in such a way that they have some relationship to each other, and establish if the relationship of those motifs exceeds the complexity threshold for a biological source. Greater specificity can provide a higher bound (e.g. specifying the type of each bone in a skeleton, rather than labelling each vertex of the graph as a bone, will result in a higher complexity value).

With this approach, we can examine complex patterns within non-identical structures comprised of non-identical parts. As an extreme example, if we were to find sets of entirely different objects (pebbles, bits of wood, etc.) joined by lengths of string on a beach, we could then construct $G$ using "any object" as a vertex and "joined by string" as an edge. If the objects were all joined in pairs, then $G$ would be simple and indeed one could imagine plausible physical effects for such a phenomenon. However, if $G$ were particularly complex and abundant, i.e. the same complex pattern were found in multiple locations, one would have to consider that some biological agency was involved. Note here that characteristics such as the lengths of the string or the shapes of the object are not considered – the connected structures could be entirely different sizes and made of completely different things, but the identical complex connectivity motif common to all of them would be enough to make a judgement on the probability of naturally occurring origin from that perspective alone.

**Finding threshold between non biological and biological systems**

In order to assess a reasonable threshold for a given set of objects, we can examine the likelihood of objects of varying complexity being constructed randomly [28]. For example, if we examine a large random text string, and look at the abundance of repeating fragments up to a certain size, we can get some idea of how the abundance of repeated fragments of increasing size drops off as the size increases. To illustrate this point, we have generated a random string of 100,000 characters and plotted the number of repeats of string fragments of different sizes up to size 8 (see Figure 6). We can see here in plot 1 that the number of repeated units drops off dramatically (note that the y-axis is logarithmic in plots 1 and 2), with very few repeats above length 4 or 5. By splicing in the word "complex" 1000 times at random positions (plot 2) we dramatically increase the number of repeating units at larger sizes. The difference in the number of repeats can be seen in plot 3, with a large difference starting at size 4. From this we can tell that we would expect to find a rather large number of repeats of size 2 and 3, but finding any abundance of repeated strings of size 7 or 8 suggests some internal structure. We can then set a reasonable threshold at 3 to 5 sigma higher than the random data, suggesting that if we find an abundance of repeats of greater sizes we have a biosignature.

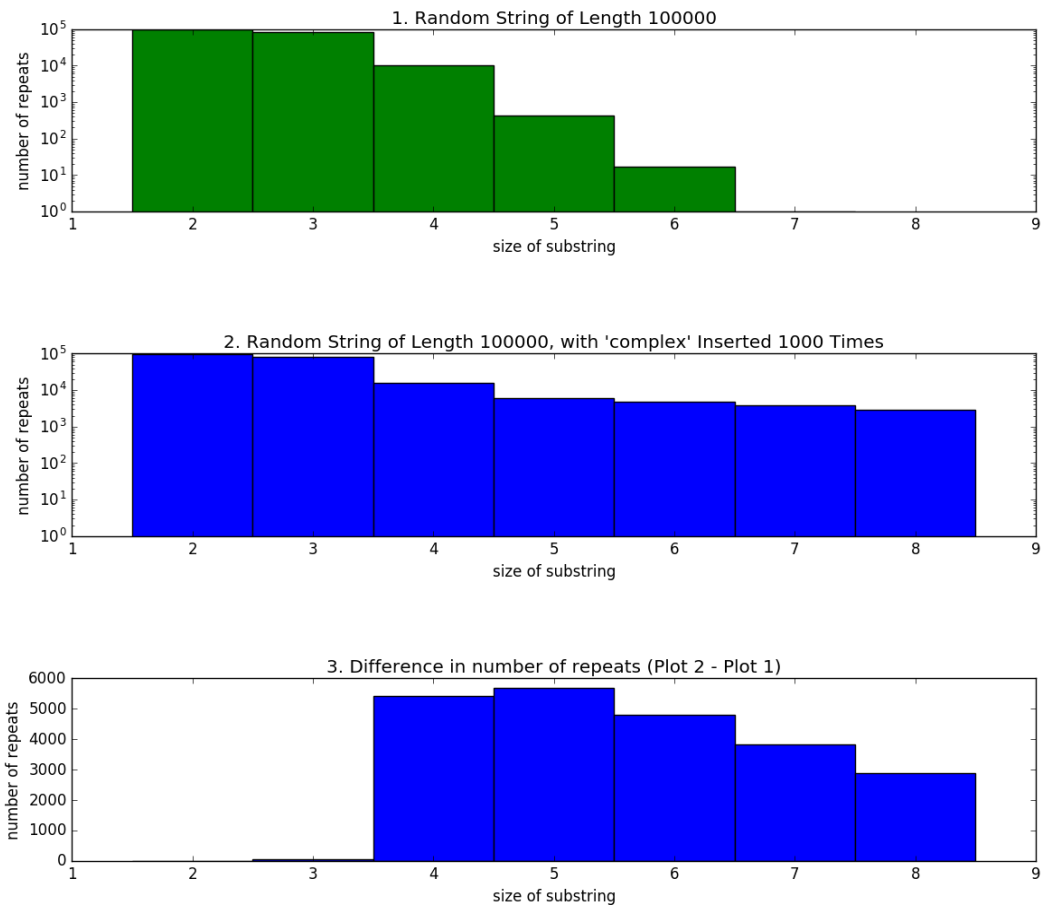

*Figure 6: Plot 1 (top) – number of repeated substrings of each size in a 100,000 character random text string. Plot 2 (middle) – as in plot 1 but where the same random string has the word 'complex' inserted at random locations 1000 times. Plot 3 (bottom) – The difference in the number of repeated substrings between Plot 2 and Plot 1.*

Although a useful indicator, this thresholding exercise does not capture the full details of the complexity measure. A stochastic model is currently in development to give a more accurate assessment of populations of larger structures being found amongst all the structures that could be created in the same time period. This will be then developed in synergy with an experimental system to investigate if a given dataset contains a biosignature or not.

**Variant: Recursive Tree Complexity**

A variant on the concept of Pathway Complexity as described above is what we have called "Recursive Tree Complexity". In this variant, we establish a complexity pathway by partitioning the object graph into a number of different subgraphs. Then the complexity of that pathway is established as the complexity of each unique subgraph, plus the number of times it is duplicated. If the subgraph is a single vertex then it contributes 1 to the complexity. The procedure is repeated recursively on the unique subgraphs, while adding 1 complexity for each time they are duplicated, and the entire structure will eventually be broken down to single-vertex graphs. The partitioning which gives the lowest total complexity is defined as the Recursive Tree Complexity. The Recursive Tree variant provides a slightly different model of the natural construction of objects. In this variant, the parts that have come together to make a particular substructure cannot be leveraged to create multiple completely different structures. In the Recursive Tree variant, one can think of different structures developing separately and then being brought together, rather than all structures being available at all times in the one pool. Any pathway in the Recursive Tree variant can also be made by the Pathway Complexity process, but it may not be the shortest pathway and may include redundant steps. Thus Recursive Tree Complexity is an upper bound for Pathway Complexity. Note also that since in Pathway Complexity the first step is a joining step, but in the Recursive Tree variant we effectively lay down a single fundamental structure first, the equivalent pathways in the Recursive Tree variant will be 1 greater than in the Pathway Complexity measure, and this must be accounted for when comparing the two.

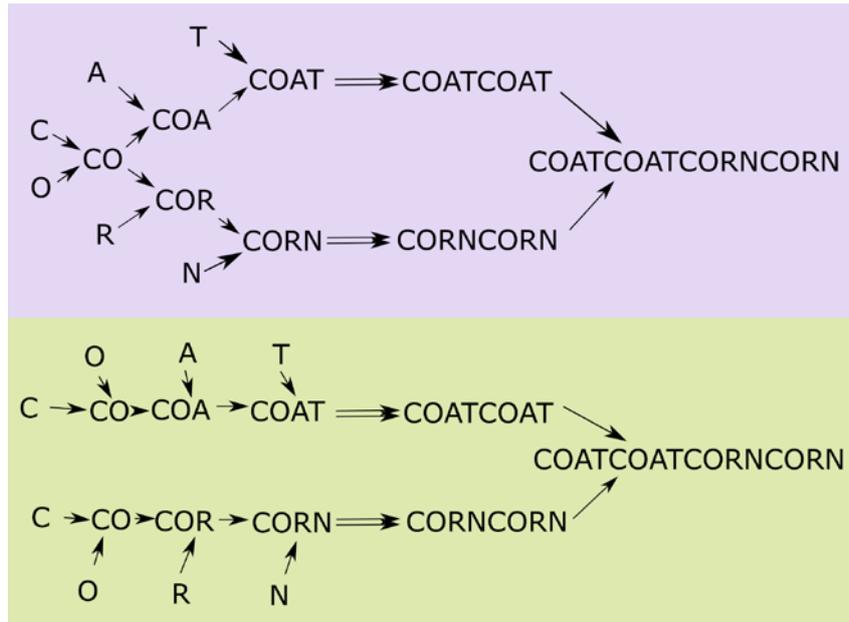

*Figure 7: The Pathway Complexity measure (top) and Recursive Tree variant (bottom) are used to construct the phrase "COATCOATCORNCORN".*

Figure 7 illustrates an example of the difference between Pathway Complexity and the Recursive Tree variant. In the former (top of the figure), the substring "CO" can be constructed and then used to make both "COA" and "COR". In the Recursive Tree variant, such sharing is not allowed, and the "CO" that goes to make "COA" is constructed separately from that which goes to make "COR".

A general mathematical formulation of the Recursive Tree variant is given below:

A **single complexity measure** $c_P$ of an object partitioned into a given multiset P is

$$c_P = \sum_{i=1}^{|K|} C_{K_i} + D(P)$$

Where K is the set of unique objects in P, $C_{K_i}$ is the Recursive Tree complexity of the $i^{\text{th}}$ member of K, and $D(P)$ is a function of the multiplicity of the objects in P (initially, $D(P)$ is the total number of duplicated objects, or formally $D(P) = \sum_{i=1}^{|P|}(|P_i| - 1)$).

The **recursive tree complexity** $C$ of an object is equal to 1 if the object cannot be partitioned further (it is atomic / its graph is a single vertex), otherwise it is equal to the minimum single complexity measure

$$C = \begin{cases} 1, & |P| = 1 \ \forall \ P \\ \min c_P, & \text{otherwise} \end{cases}$$

**Conclusions**

It is clear that biological, and biologically derived systems have an ability to create complex structures, whether proteins or iPhones, that is not found elsewhere in nature. Assessing the complexity of an object in such a way that we can define a threshold above which biology is required could be used as a biosignature in the search for shadow biologies on Earth [29], or life elsewhere in the Solar System, and would make no assumptions about the details of the biology found. We propose Pathway Complexity as a potential measure for both assessing this threshold, and determining whether objects lie above it and are therefore biologically derived. This approach provides a probabilistic context to the extending the physical basis for life detection proposed by Lovelock [30]. In further work we will show how this applies to a range of other systems, and propose a series of experimental approaches to the detection of objects and data that could be investigated as a possible biosignature. In the laboratory, we are interested in using this approach to develop a system that can explore the threshold between a non-living and living system, but also to allow us to develop a new theory for biology. This might inform a new way to search for life in the lab in terms of the complex products a system produces and if they could have arisen in any abundance by chance, rather than trying to measure the intrinsic complexity of the living system itself.

**Acknowledgements**
We would like to thank Dr Alon Henson for useful discussions.

**Funding statement**
We gratefully acknowledge financial support from the EPSRC for funding (grants EP/P00153X/1; EP/L0236521/1; EP/J015156/1), The John Templeton Foundation Grant ID